%% file: MarWai06c_arxiv.tex
\documentclass[10pt,conference]{IEEEtran}

\usepackage{amsmath,amssymb,graphics,amsthm,psfrag,subfigure}
\usepackage{amsfonts,epsfig,graphicx,subfigure}
\usepackage{amsmath,amssymb,amsthm}
\usepackage{epic,eepic}
\usepackage{epsf}
\usepackage{fancyheadings}
\usepackage{graphics}
\usepackage{amsfonts}
\usepackage{amsmath}
\usepackage{amssymb}
\usepackage{epic}
\usepackage{eepic}
\usepackage{eepicemu}
\usepackage{psfrag}

\input{defs}
\input{math_defs}

\input{rv_defs}

\input{generic_defs}

\input{variable_defs}

\input{paper_defs}
\input{martin_macros}

\newlength{\cheatLength}
\newcommand{\mypara}[1]{\noindent {\bf{#1}}}
\newcommand{\myemph}[1]{\noindent {\emph{#1}}}

\makeatletter
\long\def\@makecaption#1#2{
        \vskip 0.8ex
        \setbox\@tempboxa\hbox{\small {\bf #1:} #2}
        \parindent 1.5em  
        \dimen0=\hsize
        \advance\dimen0 by -3em
        \ifdim \wd\@tempboxa >\dimen0
                \hbox to \hsize{
                        \parindent 0em
                        \hfil 
                        \parbox{\dimen0}{\def\baselinestretch{0.96}\small
                                {\bf #1.} #2
                                } 
                        \hfil}
        \else \hbox to \hsize{\hfil \box\@tempboxa \hfil}
        \fi
        }
\makeatother

\makeatletter
\long\def\barenote#1{
    \insert\footins{\footnotesize
    \interlinepenalty\interfootnotelinepenalty 
    \splittopskip\footnotesep
    \splitmaxdepth \dp\strutbox \floatingpenalty \@MM
    \hsize\columnwidth \@parboxrestore
    {\rule{\z@}{\footnotesep}\ignorespaces
      #1\strut}}}
\makeatother

\newcommand{\midbit}{\ensuremath{m}}

\newcommand{\myber}{\ensuremath{\operatorname{Ber}}}
\newcommand{\rateclass}{\ensuremath{R}}
\newcommand{\ratewz}{\ensuremath{R_{\operatorname{WZ}}}}

\newcommand{\rateie}{\ensuremath{R_{\operatorname{IE}}}}
\newcommand{\bernoise}{\ensuremath{\delta}}

\newcommand{\rvsW}{\ensuremath{\genericRVS{w}}}
\newcommand{\rvsY}{\ensuremath{\genericRVS{y}}}

\newcommand{\sS}{\ensuremath{\genericS{s}}}
\newcommand{\sE}{\ensuremath{\genericS{e}}}

\newcommand{\sM}{\ensuremath{\genericS{m}}}
\newcommand{\sY}{\ensuremath{\genericS{y}}}
\newcommand{\sX}{\ensuremath{\genericS{x}}}
\newcommand{\sZ}{\ensuremath{\genericS{z}}}
\newcommand{\sU}{\ensuremath{\genericS{u}}}

\newcommand{\sMid}{\ensuremath{\genericS{z}}}

\newcommand{\sNoise}{\ensuremath{\genericS{v}}}

\newcommand{\topbit}{\ensuremath{n}}
\newcommand{\lowbit}{\ensuremath{k}}
\newcommand{\mydefn}{\ensuremath{: \, =}}
\newcommand{\estim}[1]{\ensuremath{\widehat{#1}}}

\newcommand{\iew}{\ensuremath{w}}
\newcommand{\channoise}{\ensuremath{p}}

\newcommand{\uce}{\ensuremath{\operatorname{u.c.e.}}}
\newcommand{\lce}{\ensuremath{\operatorname{l.c.e.}}}

\newcommand{\rateeff}{\ensuremath{R_{\operatorname{trans}}}}

\newcommand{\zeroes}{\ensuremath{0^\topbit}}

\begin{document}

\title{Low-density constructions can achieve the Wyner-Ziv and
Gelfand-Pinsker bounds }

\author{\authorblockN{Emin Martinian}
\authorblockA{Mitsubishi Electric Research Labs \\
Cambridge, MA  02139, USA \\
Email: martinian@merl.com}
\and
\authorblockN{Martin J. Wainwright}
\authorblockA{Dept. of Statistics and Dept. of EECS, \\
UC Berkeley,
Berkeley, CA  94720\\
Email: wainwrig@$\{$eecs,stat$\}$.berkeley.edu}}

\maketitle

\begin{abstract}
We describe and analyze sparse graphical code constructions for the
problems of source coding with decoder side information (the Wyner-Ziv
problem), and channel coding with encoder side information (the
Gelfand-Pinsker problem).  Our approach relies on a combination of
low-density parity check (LDPC) codes and low-density generator matrix
(LDGM) codes, and produces sparse constructions that are
simultaneously good as both source and channel codes.  In particular,
we prove that under maximum likelihood encoding/decoding, there exist
low-density codes (i.e., with finite degrees) from our constructions
that can saturate both the Wyner-Ziv and Gelfand-Pinsker bounds.
\end{abstract}

\renewcommand{\thefootnote}{\fnsymbol{footnote}}
\footnotetext{EM was supported by Mitsubishi Electric Research Labs and MJW
was supported by an Alfred P. Sloan Foundation Fellowship, an Okawa
Foundation Research Grant, and NSF Grant DMS-0528488.}

\renewcommand{\thefootnote}{\arabic{footnote}}

\section{Introduction}

Sparse graphical codes, particularly low-density parity check (LDPC)
codes, are widely used and well understood in application to channel
coding problems~\cite{Richardson:it:2001}.  For other communication
problems, especially those involving aspects of both channel and
source coding, there remain various open questions associated with
using low-density code constructions.  Two important examples are
source coding with side information (the Wyner-Ziv problem), and
channel coding with side information (the Gelfand-Pinsker problem).
This paper focuses on the design and analysis of low-density
codes---more specifically, constructions based on a combination of
LDPC and low-density generator matrix (LDGM) codes---for source and
channel coding with side information.  It builds on our previous
work~\cite{MarWai06a}, in which we proved that low-density
constructions and ML decoding can saturate the rate-distortion bound
for a symmetric Bernoulli source. \\

\mypara{Related work:} It is well-known that random constructions of
nested codes can saturate the Wyner-Ziv and Gelfand-Pinsker
bounds~\cite{WynerZiv76,Zamir02}.  However, an unconstrained random
construction leads to a high-density code, which is of limited
practical use.  One practically viable approach to lossy compression
is trellis coded quantization (TCQ)~\cite{Marcellin90}.  A number of
researchers have exploited TCQ as a quantizer for the Wyner-Ziv and
related multiterminal source coding problems~\cite{Chou03,Yang05} as
well as for channel coding with side information~\cite{}.  A
disadvantage of TCQ is that saturating rate-distortion bounds requires
that the trellis constraint length be taken infinity~\cite{Viterbi74};
consequently, the computational complexity of decoding, even using
message-passing algorithms, grows exponentially.  It is therefore of
considerable interest to develop low-density graphical constructions
for such problems.  Past work by a number of
researchers~\cite{martinian:2003:allerton,wainwright:2005:isit,Ciliberti05b,Murayama04}
has suggested that LDGM codes, which arise as the duals of LDPC codes,
are well-suited to various types of quantization.  \\

\mypara{Our contributions:} In this paper, we describe a sparse
graphical construction for generating nested codes that are
simultaneously good as both source and channel codes.  We build on our
previous work~\cite{MarWai06a}, in which we analyzed constructions,
based on a combination of LDPC and LDGM codes, for the problem of
standard lossy compression.  Here we prove that there exist variants
of these joint LDPC/LDGM constructions with finite degrees such that,
when decoded/encoded using maximum likelihood, can saturate the
Wyner-Ziv and Gelfand-Pinsker bounds.  Although ML decoding is not
practically viable, the low-density nature of our construction means
that they have low degree, and with high probability (w.h.p.) high
girth and expansion, all of which are important for the application of
efficient message-passing.

The remainder of this paper is organized as follows.
Section~\ref{SecBackground} provides background on source coding with
side information (SCSI, or the Wyner-Ziv problem), and channel coding
with side information (CSSI, or the Gelfand-Pinsker
problem). Section~\ref{SecGenConstruc} introduces our joint LDGM/LDPC
construction, and provides a high-level overview of its use for the
SCSI and CCSI problems.  In Section~\ref{SecTheory}, we prove that our
construction produces codes that are simultaneously ``good'' for both
source and channel coding.  We conclude with a discussion in
Section~\ref{SecDiscussion}. \\

\mypara{Notation:} Vectors/sequences are denoted in bold (\eg,
$\sSrc$), random variables in sans serif font (\eg, $\rvSrc$), and
random vectors/sequences in bold sans serif (\eg, $\rvsSrc$).
Similarly, matrixes are denoted using bold capital letters (\eg,
$\genMat$) and random matrixes with bold sans serif capitals (\eg,
$\rvGenMat$).  We use $I(\cdot;\cdot)$, $H(\cdot)$, and
$\rent{\cdot}{\cdot}$ to denote mutual information, entropy, and
relative entropy (Kullback-Leibler distance), respectively.  Finally,
we use $\cardinal{\{\cdot\}}$ to denote the cardinality of a set,
$\pNorm{\cdot}{p}$ to denote the $p$-norm of a vector, $\myber(t)$ to
denote a Bernoulli-$t$ distribution, and $\binent{t}$ to denote the
entropy of a $\myber(t)$ random variable.

\section{Background}
\label{SecBackground}

\subsection{Source and channel coding}

We begin with definitions of ``good'' source and channel codes that
are useful for future reference.

\begin{defn} 
(a) A code family is a \emph{good $\distor$-distortion binary
symmetric source code} if for any $\epsilon > 0$, there exists a code
with rate $\rate < 1 - \binent{\distor} + \epsilon$ that achieves
distortion $\distor$. \\
(b) A code family is a \emph{good BSC($\channoise$)-noise channel
code} if for any $\epsilon > 0$ there exists a code with rate $\rate >
1 - \binent{\channoise} - \epsilon$ with error probability less than
$\epsilon$.
\end{defn}

\subsection{Wyner-Ziv problem}
Suppose that we wish to compress a symmetric Bernoulli source $\rvsSrc
\sim \myber(\frac{1}{2})$ so as to be able to reconstruct it with
Hamming distortion $\distor$.  By classical rate distortion
theory~\cite{cover91:book}, the minimum achievable rate is given by
$\rateclass(\distor) = 1 - \binent{\distor}$.  In the Wyner-Ziv
extension~\cite{WynerZiv76}, there is an additional source of side
information about $\rvsSrc$---say in the form \mbox{$\rvsY = \rvsSrc
\oplus \rvsW$} where $\rvsW \sim \myber(\bernoise)$ is observation
noise---that is available only at the decoder.  In this setting, the
minimum achievable rate takes the form \mbox{$\ratewz(\distor,
\channoise) = \lce \big \{ \binent{\distor \ast \channoise} -
\binent{\distor}, \, (\channoise,0) \big \}$,} where $\lce$ denotes
the lower convex envelope.  Note that in the special case $\channoise
= \frac{1}{2}$, the side information is useless, so that the Wyner-Ziv
rate reduces to classical rate-distortion.
\subsection{Gelfand-Pinkser problem}

Now consider the binary information embedding problem: the channel has
the form $\sY = \sU \oplus \sS \oplus \sZ$, where $\sU$ is the channel
input, $\sS$ is a host signal (not under control of the encoder), and
$\sZ \sim \myber(\channoise)$ is channel noise.  The encoder is free
to choose the input vector $\sU \in \{0,1\}^\topbit$, subject to the
channel constraint $\|\sU \|_1 \leq \iew \topbit$, so as to maximize
the rate of information transfer.  We write $\sU \equiv \sU_\sM$ where
$\sM$ is the underlying message to be transmitted.  The decoder wishes
to recover the embedded message from the corrupted observation $\sY$.
It can be shown~\cite{Barron:2003:it} that the capacity in this set-up
is given by \mbox{$\rateie(\iew, \channoise) \uce \big\{ \binent{\iew}
- \binent{\channoise}, (0,0) \big \}$,} where $\uce$ denotes the upper
convex envelope.

\section{Generalized Compound Construction}
\label{SecGenConstruc}

In this section, we describe a compound construction that produces
codes that are simultaneously ``good'', in the senses previously
defined, as source and channel codes.  We then describe how the nested
codes generated by this compound construction apply to the SCSI and
CCSI problems.

\subsection{Code construction}

\begin{figure}[h]
\begin{center}
\psfrag{#k#}{$\lowbit$} \psfrag{#k1#}{$\lowbit_1$}
\psfrag{#k2#}{$\lowbit_2$} \psfrag{#m#}{$\midbit$}
\psfrag{#topdeg#}{$\topdeg$} \psfrag{#vdeg#}{$\vdeg$}
\psfrag{#cdeg#}{$\cdeg$} \psfrag{#n#}{$n$} \psfrag{#G#}{$\genMat$}
\psfrag{#H1#}{$\parMat_1$} \psfrag{#H2#}{$\parMat_2$}
\widgraph{.51\textwidth}{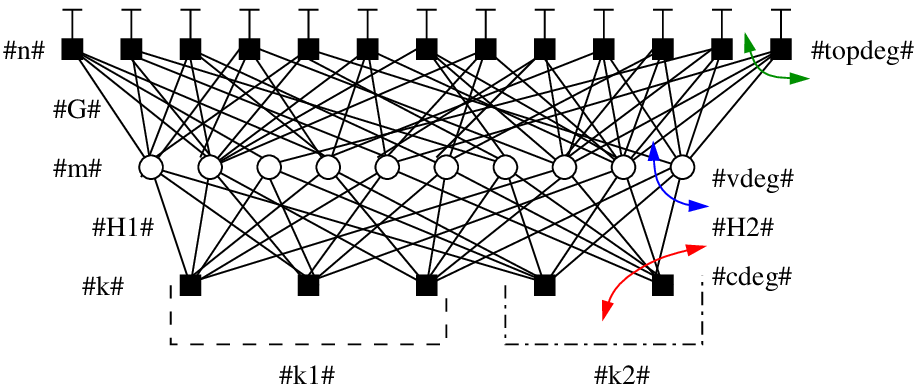}
\caption{Illustration of compound LDGM and LDPC code construction.
The top section consists of an $(\topbit, \midbit)$ LDGM code with
generator matrix $\genMat$ and constant check degrees $\topdeg = 4$;
its rate is $\rateldgm = \frac{\midbit}{\topbit}$.  The bottom section
consists of $(\midbit, \lowbit_1)$ and $(\midbit, \lowbit_2)$ LDPC
codes with degrees $(\vdeg, \cdeg) = (3,6)$, described by parity check
matrices $\parMat_1$ and $\parMat_2$ and rates $\rateone =
1-\frac{\lowbit_1}{\midbit}$ and $\ratetwo =
1-\frac{\lowbit_2}{\midbit}$ respectively.  The overall rate of the
compound construction is \mbox{$\ratetot = \rateldgm \rateldpc)$,}
where $\rateldpc = \rateone + \ratetwo$.}
\label{FigGenCompound}
\end{center}
\vspace{-\cheatLength}
\end{figure}

Consider the compound code construction illustrated
in~\figref{FigGenCompound}, defined by a factor graph with three
layers.  The top layer consists of $\topbit$ bits, each attached to an
associated parity check.  These parity checks connect to $\midbit$
variable nodes in the middle layer, and in turn these middle variable
nodes are connected to $\lowbit = \lowbit_1 + \lowbit_2$ parity checks
in the bottom layer.

\myemph{Random LDGM ensemble:} The top two layers define an $(\topbit,
\midbit)$ LDGM code.  We construct it by connecting each of the
$\nbit$ checks at the top randomly to $\topdeg$ variable nodes in the
middle layer chosen uniformly at random.  We use $\genMat \in
\{0,1\}^{\midbit \times \topbit}$ to denote the resulting generator
matrix; by construction, each column of $\genMat$ has exactly
$\topdeg$ ones, whereas each row (corresponding to a variable node)
has an (approximately) Poisson number of ones.  An advantage of this
regular-Poisson degree ensemble is that the resulting distribution of
a random codeword is extremely easy to characterize:
\begin{lems}
\label{LemInducedDmin}
Let $\rvGenMat \in \{0,1\}^{\midbit \times \topbit}$ be a random
generator matrix obtained by randomly placing $\topdeg$ ones per
column.  Then for any vector $\sIntm \in \{0,1\}^\midbit$ with a
fraction of $\weight$ ones, the distribution of the corresponding
codeword $\sIntm \, \rvGenMat$ is
Bernoulli($\inducedWeight{\weight;\topdeg}$) where
\begin{equation}
\label{eq:InducedWeight}
\inducedWeight{\weight;\topdeg} = \frac{1}{2} \cdot \left[1 - (1-2
  \weight)^{\topdeg}\right].
\end{equation}
\end{lems}

\myemph{Random LDPC code:} The bottom two layers define a pair of LDPC
codes, with parameters $(\midbit, \lowbit_1)$ and $(\midbit,
\lowbit_2)$; we choose these codes from a standard standard $(\vdeg,
\cdeg)$-regular LDPC ensemble originally studied by Gallager.
Specifically, each of the $\midbit$ variable nodes in the middle layer
connects to $\vdeg$ check nodes in the bottom layer.  Similarly, each
of the $\lowbit$ check nodes in the bottom layer connects to $\cdeg$
variable nodes in the middle layer.  For convenience, we restrict
ourselves to even check degrees $\cdeg$.  Dividing the $\lowbit$ check
bits into two subsets, of size $\lowbit_1$ and $\lowbit_2$ with
respective parity check matrices $\parMat_1$ and $\parMat_2$, allows
for the construction of nested codes, which will be needed for both
the Wyner-Ziv and Gelfand-Pinsker problems.

\subsection{Good source and channel codes}

The key theoretical properties of this joint LDGM/LDPC construction
are summarized in the following results:
\begin{theos}[Good source code] 
\label{ThmSource}
With appropriate finite degrees, there exist $(\topbit, \midbit,
\lowbit)$ constructions that are $\distor$-good source codes for all
rates above $R(\distor) = 1 - \binent{\distor}$.
\end{theos}

\begin{theos}[Good channel code] 
\label{ThmChannel}
With appropriate finite degrees, there exist $(\topbit, \midbit,
\lowbit)$ constructions that are good $\channoise$-channel codes for
all rates below capacity $C = 1 - \binent{\channoise}$.
\end{theos}
Theorem~\ref{ThmSource} on source coding was proved in our previous
work \cite{MarWai06a}, whereas a proof of Theorem~\ref{ThmChannel} is
given in Section~\ref{SecTheory}.  We now describe how these two
theorems allow us to establish that our low-density construction
achieves the Wyner-Ziv and Gelfand-Pinsker bounds.  At a high level,
our approach is closely related to standard approaches to SCSI/CSCI
coding; the key novelty is that appropriately nested codes can be
construction using low-density architectures.

\subsection{Coding for Wyner-Ziv}

We focus only on achieving rates of the form $\binent{\distor \ast
  \channoise} - \binent{\distor}$, as any remaining rates on the
Wyner-Ziv curve can be achieved by time-sharing with the point
$(\channoise, 0)$.  To do this, we use the compound code
in \figref{FigGenCompound}.  Specifically, a source $\rvsSrc$ is
encoded to $\parMat_2\sIntm$ where $\sIntm$ is chosen to minimize the
distortion $\pNorm{\rvsSrc-\sIntm'\genMat}{1}$ subject to the constraint
that $\parMat_1\sIntm=0$.  Theorems~\ref{ThmSource}
and~\ref{ThmChannel} show that maximum likelihood decoding of
$\parMat_2\sIntm$ using side information $\rvsY$ approaches the
Wyner-Ziv bound in the sense that this construction yields a
good $\distor$-distortion binary source code, and a nested subcode that
is a good $D \ast \channoise$-noise channel code.  Details follow.

\mypara{Source coding component:} The $\distor-$distortion source code
component of the construction involves the $\topbit$ variable nodes
representing the source bits, the $\midbit$ intermediate variable
nodes, and the subset of $\lowbit_1$ lower layer check nodes.  This
subgraph, represented by the generator matrix $\genMat$ and parity
check matrix $\parMat_1$ (see~\figref{FigGenCompound}), define
a code (on the $\topbit$ variable nodes) with effective rate
\begin{equation}
\rate_1 \mydefn \frac{\midbit \, \big(1 -
\frac{\lowbit_1}{\midbit}\big)}{\nbit} \; = \;
\frac{\midbit-\lowbit_1}{\nbit}.
\end{equation}
Choosing the middle and lower layer sizes $\midbit$ and $\lowbit_1$
such that $\rate_1 = 1 - \binent{\distor}$ guarantees (from
Theorem~\ref{ThmSource}) the existence of finite degrees such that
that this code is a good $\distor$-distortion source code.


\mypara{Channel coding component:} Now suppose that the source
$\sSrc$ has been quantized, and is represented (up to
distortion $\distor$) by the compressed sequence $\estim{\sX} \in
\{0,1\}^\midbit$.  We transmit the associated sequence $\parMat_2
\estim{\sX} \in \{0,1\}^{\lowbit_2}$ of parity bits associated with
the code $\parMat_2$; doing so requires rate $\rateeff =
\frac{\lowbit_2}{\topbit}$.  The task of the decoder is as follows:
given these $\lowbit_2$ parity bits as well as the $\lowbit_1$
zero-valued parity bits, the decoder seeks to recover the quantized
sequence $\estim{\sX}$ on the basis of the observed side-information
$\sY$.  Note that from the decoder's perspective, the effective code
rate is given by
\begin{eqnarray}
\rate_2 & = & \frac{\midbit - \lowbit_1 - \lowbit_2}{\topbit}
\end{eqnarray}
Suppose that we choose $\lowbit_2$ such that $\rate_2 = 1 -
\binent{\distor \ast \channoise}$; then Theorem~\ref{ThmChannel}
guarantees that the decoder will (w.h.p.) be able to recover a
codeword corrupted by $(\distor \ast \channoise)$-Bernoulli noise.
Note that the side information can be written as \mbox{$\sY =
\estim{\sS} \oplus \sE \oplus \sNoise,$} where $\sE \mydefn \sS \oplus
\estim{\sS}$ is the quantization noise, and $\sNoise \sim
\myber(\channoise)$ is the channel noise.  If the quantization noise
$\sE$ were i.i.d. $\myber(\distor)$, then the overall effective noise
$\sE \oplus \sNoise$ would be i.i.d. $\myber(\distor \ast
\channoise)$.  In reality, the quantization noise is not exactly
i.i.d. $\myber(\distor)$, but it can be shown~\cite{Zamir02} that it
can be treated as such for theoretical purposes.

In summary then, the overall transmission rate of this scheme for the
Wyner-Ziv problem is given by
\begin{eqnarray}
\left(\frac{\midbit - \lowbit_1}{\topbit}\right) - \left(\frac{\midbit
  - \lowbit_1 - \lowbit_2}{\topbit} \right) & = & \binent{\distor \ast
  \channoise} - \binent{\distor}. \qquad
\end{eqnarray}
Thus, by applying Theorems~\ref{ThmSource} and~\ref{ThmChannel}, we
conclude that our low-density scheme saturates the Wyner-Ziv bound.

\subsection{Coding for Gelfand-Pinsker}

The construction for the Gelfand-Pinsker problem is similar, but
with the order of the code nesting reversed.  In particular, the
Gelfand-Pinsker problem requires a good $\channoise$-noise channel
code, and a nested subcode that is a good $\iew$-distortion source
code.  As before, we focus only on achieving rates of the form
$\binent{\iew} - \binent{\channoise}$.  To encode a message $\sM$ with
side information $\rvsY$, the channel input is $\sIntm'\genMat$ where
$\sIntm$ is chosen to minimize $\pNorm{\rvsY-\sIntm'\genMat}{1}$ subject
to $\parMat_1\sIntm=\sM$.  Details follow.

\mypara{Source coding component:} We begin by describing the nested
subcode for the source coding component.  The idea is to embed a
message into the transmitted signal during the quantization process.
The first set of $\lowbit_1$ lower parity bits remain fixed to zero
throughout the scheme.  On the other hand, we use the remaining
$\lowbit_2$ lower parity bits to specify a particular message $\sM \in
\{0,1\}^{\lowbit_2}$ that the decoder would like to recover.  With the
lower parity bits specified in this way, we use the resulting code to
quantize a given source sequence $\sS$ to a compressed version
$\estim{\sS}$.  If we choose $\topbit, \midbit$ and $\lowbit$
such that
\begin{eqnarray}
\rate_1 & = & \frac{\midbit - \lowbit_1 - \lowbit_2}{\topbit} \; = \;
1 - \binent{\iew},
\end{eqnarray}
then Theorem~\ref{ThmSource} guarantees that the resulting code is a
good $\iew$-distortion source code.  Otherwise stated, we are
guaranteed that w.h.p, the error $\sE \mydefn \sS \oplus \estim{\sS}$
in our quantization has Hamming weight upper bounded by $\iew
\topbit$.  Thus, transmitting the error $\sE$ ensures that the channel
constraint is met.

\mypara{Channel coding component:} At the decoder, the $\lowbit_1$
lower parity bits remain set to zero; the remaining $\lowbit_2$ parity
bits, which represent the message $\sM$, are unknown to the coder.  We
now choose $\lowbit_1$ such that the effective code used by the
decoder has rate
\begin{eqnarray} 
\label{EqnChanRateGelfand}
\rate_2 & = & \frac{\midbit - \lowbit_1}{\topbit} = 1 -
\binent{\channoise}.
\end{eqnarray}
In addition, the decoder is given a noisy channel observation of the
form \mbox{$\sY = \sE \oplus \sS \oplus \sNoise \; = \; \estim{\sS}
\oplus \sNoise$} and its task is to recover $\estim{\sS}$.  With the
channel coding rate chosen as in equation~\eqref{EqnChanRateGelfand}
and channel noise $\sNoise \sim \myber(\channoise)$,
Theorem~\ref{ThmChannel} guarantees that the decoder will w.h.p. be
able to recover $\estim{\sS}$.  By design of the quantization
procedure, we have the equivalence $\sM = \estim{\sS} \, \parMat_2$ so
that a simple syndrome-forming procedure allows the decoder to recover
the hidden message.  Thus, by applying Theorems~\ref{ThmSource}
and~\ref{ThmChannel}, we conclude that our low-density scheme
saturates the Gelfand-Pinsker bound under ML encoding/decoding.


\section{Proof of Theorem~\ref{ThmChannel}}
\label{SecTheory}

As described in the previous sections, Theorems~\ref{ThmSource}
and~\ref{ThmChannel} allow us to establish that the Wyner-Ziv and
Gelfand-Pinsker bounds can be saturated under ML encoding/decoding.
The source coding part---namely Theorem~\ref{ThmSource}---was proved
in our earlier work~\cite{MarWai06a}.  Here we provide a proof of
Theorem~\ref{ThmChannel}, which ensures that these joint LDGM/LDPC
constructions are good channel codes.  We consider a joint
construction, as illustrated in~\figref{FigGenCompound}, consisting of
a rate $\rateldgm$ LDGM top code, and a rate $\rateldpc$ lower LDPC
code.  Recall that the overall rate of this compound construction is
given by $\ratetot = \rateldgm \rateldpc$.  Note that an LDGM code on
its own (i.e., without the lower LDPC code) is a special case of this
construction with $\rateldpc = 1$.  However, a standard LDGM of this
variety is \emph{not} a good channel code, due to the large number of
low-weight codewords.  Essentially, the following proof shows that
using a non-trivial LDPC lower code (with $\rateldpc < 1$) can
eliminate these troublesome low-weight codewords.

If the codeword $\cword$ is transmitted, then the receiver observes
$\sY = \cword \oplus \sNoise$ where $\sNoise$ is a
$\myber(\channoise)$ random vector.  Our goal is to bound the
probability that maximum likelihood (ML) decoding fails where the
probability is taken over the randomness in both the channel noise and
the code construction.  To simplify the analysis, we focus on the
following sub-optimal (non-ML) decoding procedure:
\begin{defn}[Decoding Rule:] With threshold $d(\topbit) \mydefn (p + \topbit^{-1/3}) \,
\topbit$, decode to codeword $\cword_i$ $\iff$ \mbox{$\|\cword_i
\oplus \sY\|_1 \leq d(\topbit)$,} and no other codeword is within
$d(\topbit)$ of $\sY$.
\end{defn}
\noindent (The extra factor of $\topbit^{-1/3}$ in the threshold
$d(\topbit)$ is of theoretical convenience.)  Due to the linearity of
the code construction, we may assume without loss of generality that
the all zeros codeword $\zeroes$ was transmitted (\ie, $\cword =
\zeroes$).  In this case, the channel output is simply $\sY = \sNoise$
and so our decoding procedure will fail if and only if either (i)
$\|\sNoise \|_1 > d(\topbit)$, or (ii) there exists some codeword
``middle layer codeword'' $\sMid \in \{0,1\}^\midbit$ satisfying the
parity check equation\footnote{To be more precise, for the channel
decoding step of the Wyner-Ziv problem, the middle layer codeword must
satisfy $\parMat_1 \, \sMid = 0$ and $\parMat_2 \, \sMid = \sM$ where
$\sM$ is the output of the Wyner-Ziv encoder.  For the channel
decoding step of the Gelfand-Pinsker problem, the middle layer
codeword must only satisfy $\parMat_1 \, \sMid = 0$, since $\sM$ is
unknown until decoding is complete.}  $\parMat \, \sMid = 0$ and
corresponding to a codeword $\cword_i = \sMid \, \genMat$ such that
$\|\sMid \, \genMat \oplus \sNoise\|_1 \leq d(\topbit)$.  Using the
following two lemmas, we establish that this procedure has arbitrarily
small probability of error, whence ML decoding (which is at least as
good) also has arbitrarily small error probability.
\begin{lems}
The probability of decoding error vanishes asymptotically provided
that
\begin{equation}
\label{EqnErrExponent}
\rateldgm \WtEnumAsymp(\weight) -
\rent{\channoise}{\inducedWeight{\weight;\topdeg} \ast \channoise} \;
< \; 0 \mbox{ for all $\weight \in (0,\myhalf]$}
\end{equation}
where $\WtEnumAsymp(\weight) \mydefn \lim_{\midbit \rightarrow
+\infty} \midWtEnum(\weight)$ is the asymptotic log-domain weight
numerator of the LDPC code, with \midWtEnum(\weight) being the average
log-domain weight enumerator defined as
\begin{eqnarray}
\midWtEnum(\weight) & \mydefn & \frac{1}{\midbit} \log \Exs
\operatorname{card} \big \{ \sMid \; \big | \; \pNorm{\sMid}{1} =
\weight \midbit \big \}.
\end{eqnarray}
\end{lems}
\begin{proof}
Let $N = 2^{\topbit \ratetot}$ denote the total number of codewords in
the joint LDGM/LDPC code.  Then we can upper bound the probability of
error using the union bound as follows:
\begin{equation}
\label{EqnErr}
p_{err} \leq \Prob [\|\sNoise\|_1 > d(\topbit) ] + \sum_{i=2}^N \Prob
\big[\|\sMid_i \, \genMat \oplus \sNoise\|_1 \leq d(\topbit) \big]. \;
\;
\end{equation}
By Bernstein's inequality, the probability of the first error event
vanishes for large $\topbit$.  Now focusing on the second sum, let us
condition on the event that $\|\sMid\|_1 = \ell$. Then
Lemma~\ref{LemInducedDmin} guarantees that $\sMid \, \genMat$ has
i.i.d. $\myber(\inducedWeight{\frac{\ell}{\midbit};\topdeg})$
elements, so that the vector $\sMid \, \genMat \oplus \sNoise$ has
i.i.d. $\myber(\inducedWeight{\frac{\ell}{\midbit};\topdeg} \ast
\channoise)$ elements.  Applying Sanov's theorem yields the upper
bound
\begin{eqnarray*}
\Prob\big[\|\sMid \, \genMat \oplus \sNoise\|_1 \geq d(\topbit) \;
\big | \; \|\sMid\|_1 = \ell \big] & \leq & 2^{-\topbit
\rent{\channoise}{\inducedWeight{\frac{\ell}{\midbit};\topdeg} \ast
\channoise}}.
\end{eqnarray*}
We can then upper bound the second error term~\eqref{EqnErr} via
\begin{align*}
& \quad 2^{\topbit \ratetot} \; \sum_{\ell=0}^\midbit \Prob[\|\sMid |_1 =
  \ell ] \; 2^{-\topbit
  \rent{\channoise}{\inducedWeight{\frac{\ell}{\midbit};\topdeg} \ast
  \channoise}} \\
&= \sum_{\ell=0}^{\midbit} 2^{ \left\{\topbit \ratetot + \midbit
 \big[\midWtEnum(\frac{\ell}{\midbit}) -\rateldpc \big] -\topbit
 \rent{\channoise}{\inducedWeight{\frac{\ell}{\midbit};\topdeg} \ast
 \channoise} \right\}} \\
&= \sum_{\ell=0}^{\midbit} 2^{ \topbit \left\{\rateldgm
\midWtEnum(\frac{\ell}{\midbit}) -
\rent{\channoise}{\inducedWeight{\frac{\ell}{\midbit};\topdeg} \ast
\channoise} \right\}}\\
&= \sum_{\ell=0}^{\midbit} 2^{ \topbit \left\{\rateldgm[
\midWtEnum(\frac{\ell}{\midbit}) -
\WtEnumAsymp(\frac{\ell}{\midbit}) +
\WtEnumAsymp(\frac{\ell}{\midbit})] 
-\rent{\channoise}{\inducedWeight{\frac{\ell}{\midbit};\topdeg} \ast
\channoise} \right\}}\\
& \leq \sum_{\ell=0}^{\midbit} 2^{ \topbit  \left\{\rateldgm
|\midWtEnum(\frac{\ell}{\midbit}) - \WtEnumAsymp(\frac{\ell}{\midbit})|^+ +
\rateldgm\WtEnumAsymp(\frac{\ell}{\midbit})
-\rent{\channoise}{\inducedWeight{\frac{\ell}{\midbit};\topdeg} \ast
\channoise} \right\}}
\end{align*}
where we have replaced $\ratetot = \rateldgm$ with $\rateldpc$ in the
third line and used the notation $|x|^+$ to denote $\max(0,x)$.
Finally, we notice that by the definition of the asymptotic weight
enumerator, $\WtEnumAsymp(\weight)$, the $|\midWtEnum(\weight) -
\WtEnumAsymp(\weight)|^+$ term converges to zero
uniformly\footnote{The definition of $\WtEnumAsymp(\weight)$ implies
  pointwise convergence of $|\midWtEnum(\weight) -
  \WtEnumAsymp(\weight)|^+$ for $\weight\in[0,1]$.  But since the
  domain is compact, pointwise convergence implies uniform convergence.}
for $\weight\in[0,1]$ leaving only the error exponent
(\ref{EqnErrExponent}), which is negative by assumption.
\end{proof}

\begin{figure*}
\begin{center}
\begin{tabular}{cc}
\widgraph{0.35\textwidth}{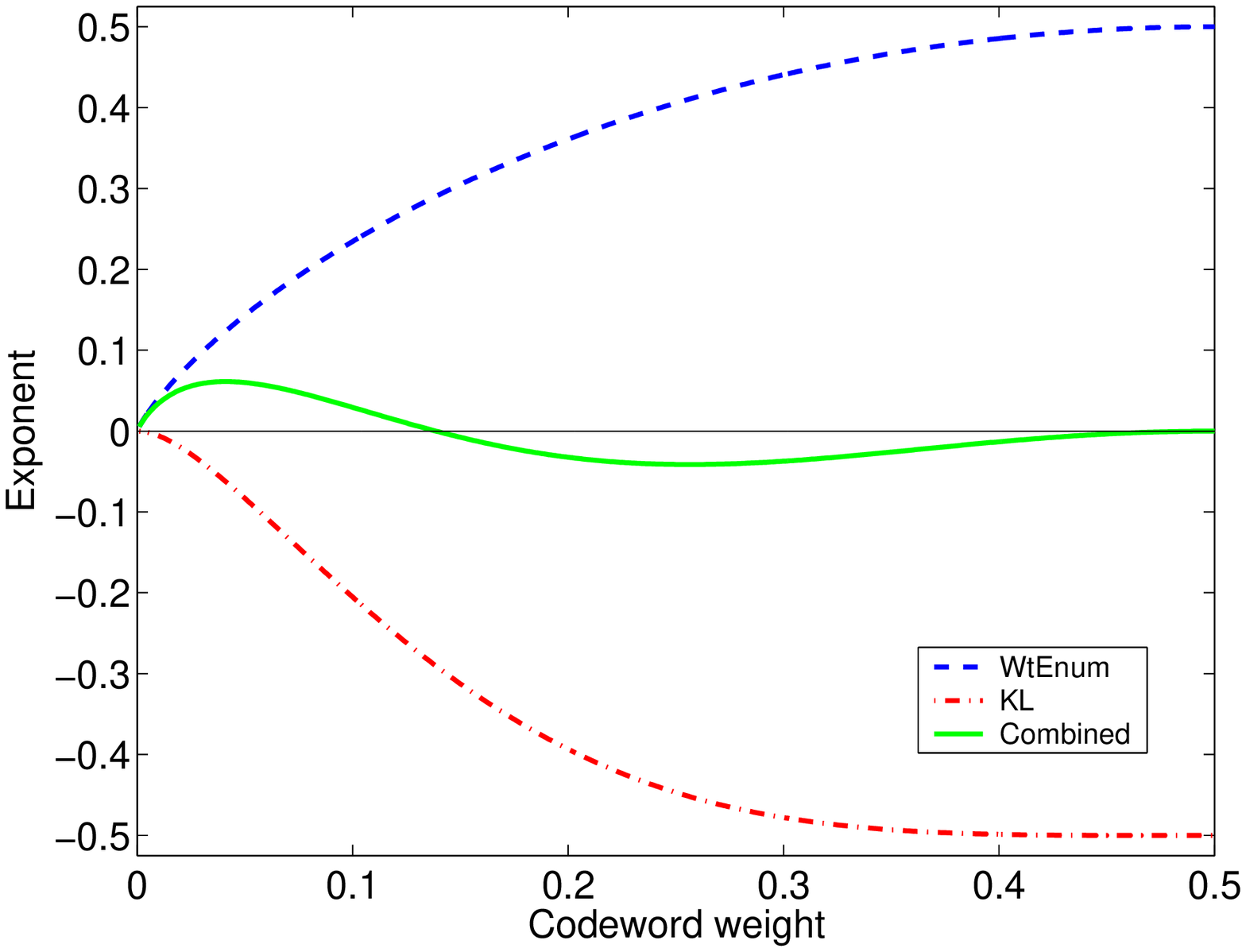} &
\widgraph{0.35\textwidth}{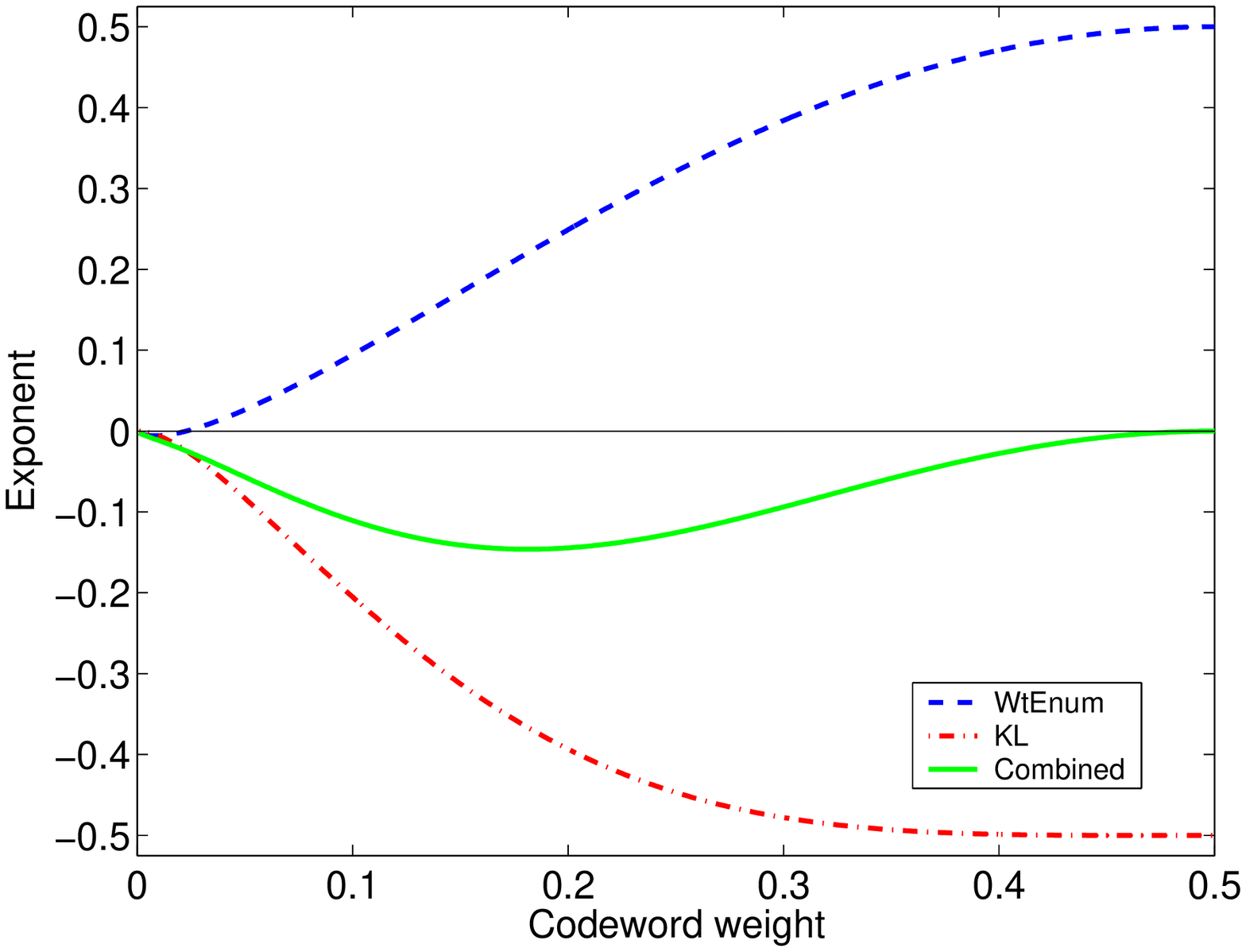} \\
(a) &
(b) 
\end{tabular}
\caption{Plots of different terms in error
exponent~\eqref{EqnErrExponent}.  The combined curve must remain
negative for all $\omega$ in order for the error probability to vanish
asymptotically.  (a) A LDGM $\topdeg=4$ construction without any LDPC
lower code: here the weight enumerator $\WtEnumAsymp$ is given by
$\binent{\omega}$, and it dominates the Kullback-Leibler term for low
$\omega$.  (b) The same $\topdeg = 4$ LDGM combined with a $(\vdeg,
\cdeg) = (3,6)$ LDPC lower code: here the LDPC weight enumerator is
dominated for all $\omega$ by the KL error exponent.}
\label{FigExponents}
\end{center}
\vspace{-\cheatLength}
\end{figure*}

\begin{lems} 
\label{lem:easy_lemma}
For any $\channoise \in (0,1)$ and total rate $\ratetot \mydefn
\rateldgm \, \rateldpc \; < \; 1 -\binent{\channoise}$, it is possible
to choose the code parameters $\topdeg$, $\cdeg$ and $\vdeg$ such that
\eqref{EqnErrExponent} is satisfied.
\end{lems}
\begin{proof}
For brevity, let \mbox{$\ErrFun(\weight) =
\rateldgm \WtEnumAsymp(\weight) -
\rent{\channoise}{\inducedWeight{\weight;\topdeg} \ast \channoise}$.}
It is well-known that a regular LDPC code with rate $\rateldpc =
\frac{\vdeg}{\cdeg} < 1$ has linear minimum distance; in particular,
there exists a threshold $\ldpcthresh = \ldpcthresh(\vdeg, \cdeg)$
such that $\WtEnumAsymp(\weight) \leq 0$ for all $\weight \in [0,
\ldpcthresh]$.  Hence, for $\weight \in (0, \ldpcthresh]$, we have
$\ErrFun(\weight) < 0$.

Turning now to the interval $[\ldpcthresh, \myhalf]$, consider the
function
\[
G(\weight) \mydefn \ratecom \binent{\weight} -
\rent{\channoise}{\inducedWeight{\weight;\topdeg}}.
\]
Since $\WtEnumAsymp(\weight) \leq \rateldpc \binent{\weight}$, we have
$\ErrFun(\weight) \leq G(\weight)$, so that it suffices to upper bound
$G$.  Observe that $G(\myhalf) = \ratecom - (1 - \binent{\channoise})
< 0$.  Therefore, it suffices to show that, by appropriate choice of
$\topdeg$, we can ensure that $G(\weight) \leq G(\myhalf)$.  Noting
that $G$ is infinitely differentiable and taking derivatives (details
omitted), it can be shown that $G'(\myhalf) = 0$ and $G''(\myhalf) <
0$.  Hence, a second order Taylor series expansion yields that
$G(\weight) \leq G(\myhalf)$ for all $\weight \in (\lowtay, \myhalf]$
for some $\lowtay < \myhalf$.  It remains to bound $G$ on the interval
$[\ldpcthresh, \lowtay]$.  On this interval, we have $G(\weight) \leq
\ratecom \binent{\lowtay} -
\rent{\channoise}{\inducedWeight{\ldpcthresh;\topdeg}}$.  By examining
\eqref{eq:InducedWeight}, we see that choosing $\topdeg$
sufficiently large will ensure that on the interval $[\ldpcthresh,
\lowtay]$, the RHS is less than $\ratecom - (1-\binent{\channoise})$
as required.
\end{proof}
\noindent Theorem~\ref{ThmChannel} follows by combining the
previous lemmas.

At first glance, Lemma~\ref{lem:easy_lemma} may seem unsatisfying,
since it might require a very large top degree $\topdeg$.  Note,
however, that this degree does not depend on the block length, hence
our claim that good low density codes can be constructed with
\emph{finite} degree.  Of course, for the claim of finite degree codes
to be practically meaningful, the degree required for $\topdeg$ should
be reasonably small.  To investigate this issue, we plot the error
exponent~\eqref{EqnErrExponent} for rate $\ratecom = 0.5$, LDGM top
degree $\topdeg = 4$, and different choices of lower code with
$\rateldpc$ in Figure~\ref{FigExponents}.  Without any lower LDPC
code, then $\rateldpc = 1$ and the effective asymptotic weight
enumerator is simply $\binent{\omega}$.  Panel (a) shows the behavior
in this case: note that the error exponent exceeds zero in a region
around $\weight = 0$ where the weight enumerator dominates the
negative KL term.  In contrast, panel (b) shows the case of a $(\vdeg,
\cdeg) = (3,6)$ LDPC code, where we have used the results of Litsyn
and Shevelev~\cite{Litsyn_and_Shevelev} in plotting the asymptotic
weight enumerator.  This code family has a linear minimum distance, so
that the log-domain weight enumerator is negative in a region around
$\weight = 0$.  Thus, the error exponent~\eqref{EqnErrExponent}
remains negative for all $\weight \in [0, 0.5]$.  Thus, provided that
a $(3,6)$ lower LDPC code is used, a very reasonable top degree of
$\topdeg = 4$ is sufficient.

\section{Discussion}
\label{SecDiscussion}

We have established that sparse graphical constructions that exploit
both LDGM and LDPC codes can saturate fundamental bounds for problems
of source coding with side information, and channel coding with side
information.  Although the present results are based on ML
encoding/decoding, the sparsity and graphical structure of our
constructions render them suitable candidates for practical
message-passing schemes, which remains to be investigated in future
work.

\bibliographystyle{plain}



\end{document}

%% file: defs.tex
\newcommand{\ie}{{\em i.e.}}

\newcommand{\eg}{{\em e.g.}}

\newcommand{\figref}[1]{Fig.~\ref{#1}}

\newtheorem{defn}{Definition}

%% file: math_defs.tex
\newcommand{\pNorm}[2]{||#1||_{#2}}

\newcommand{\cword}{\mathbf{c}}

%% file: rv_defs.tex
\DeclareMathAlphabet{\mathbsf}{OT1}{cmss}{bx}{n}
\DeclareMathAlphabet{\mathssf}{OT1}{cmss}{m}{sl}
\DeclareMathAlphabet{\mathcsf}{OT1}{cmss}{sbc}{n}

\DeclareSymbolFont{bsfletters}{OT1}{cmss}{bx}{n}  
\DeclareSymbolFont{ssfletters}{OT1}{cmss}{m}{n}
\DeclareMathSymbol{\bsfGamma}{0}{bsfletters}{'000}
\DeclareMathSymbol{\ssfGamma}{0}{ssfletters}{'000}
\DeclareMathSymbol{\bsfDelta}{0}{bsfletters}{'001}
\DeclareMathSymbol{\ssfDelta}{0}{ssfletters}{'001}
\DeclareMathSymbol{\bsfTheta}{0}{bsfletters}{'002}
\DeclareMathSymbol{\ssfTheta}{0}{ssfletters}{'002}
\DeclareMathSymbol{\bsfLambda}{0}{bsfletters}{'003}
\DeclareMathSymbol{\ssfLambda}{0}{ssfletters}{'003}
\DeclareMathSymbol{\bsfXi}{0}{bsfletters}{'004}
\DeclareMathSymbol{\ssfXi}{0}{ssfletters}{'004}
\DeclareMathSymbol{\bsfPi}{0}{bsfletters}{'005}
\DeclareMathSymbol{\ssfPi}{0}{ssfletters}{'005}
\DeclareMathSymbol{\bsfSigma}{0}{bsfletters}{'006}
\DeclareMathSymbol{\ssfSigma}{0}{ssfletters}{'006}
\DeclareMathSymbol{\bsfUpsilon}{0}{bsfletters}{'007}
\DeclareMathSymbol{\ssfUpsilon}{0}{ssfletters}{'007}
\DeclareMathSymbol{\bsfPhi}{0}{bsfletters}{'010}
\DeclareMathSymbol{\ssfPhi}{0}{ssfletters}{'010}
\DeclareMathSymbol{\bsfPsi}{0}{bsfletters}{'011}
\DeclareMathSymbol{\ssfPsi}{0}{ssfletters}{'011}
\DeclareMathSymbol{\bsfOmega}{0}{bsfletters}{'012}
\DeclareMathSymbol{\ssfOmega}{0}{ssfletters}{'012}

%% file: generic_defs.tex
\newcommand{\genericRV}[1]{\mathssf{#1}} 
\newcommand{\genericS}[1]{\mathbf{#1}} 
\newcommand{\genericRVS}[1]{\mathbsf{#1}} 
\newcommand{\genericMatrix}[1]{\mathbf{\MakeUppercase{#1}}} 
\newcommand{\genericRVMatrix}[1]{\mathbsf{\MakeUppercase{#1}}} 



%% file: variable_defs.tex
\newcommand{\src}{s} 
\newcommand{\sSrc}{\genericS{\src}} 
\newcommand{\rvSrc}{\genericRV{\src}} 
\newcommand{\rvsSrc}{\genericRVS{\src}} 

\newcommand{\intm}{w} 
\newcommand{\sIntm}{\genericS{\intm}} 


%% file: paper_defs.tex
\newcommand{\lowtay}{\ensuremath{\mu}}

\newcommand{\ErrFun}{\ensuremath{F}}

\newcommand{\ldpcthresh}{\ensuremath{\nu^*}}

\newcommand{\myhalf}{\ensuremath{\frac{1}{2}}}

\newcommand{\rateldgm}{\ensuremath{R(\genMat)}}
\newcommand{\rateldpc}{\ensuremath{R(\parMat)}}
\newcommand{\rateone}{\ensuremath{R(\parMat_1)}}
\newcommand{\ratetwo}{\ensuremath{R(\parMat_2)}}
\newcommand{\ratetot}{\ensuremath{R_{\operatorname{com}}}}
\newcommand{\ratecom}{\ensuremath{\ratetot}}

\newcommand{\binent}[1]{H_b\left(#1\right)} 
\newcommand{\rent}[2]{D\left(#1||#2\right)} 

\newcommand{\weight}{v}

\newcommand{\inducedWeight}[1]{\delta(#1)}

\newcommand{\genMat}{\genericMatrix{G}}
\newcommand{\rvGenMat}{\genericRVMatrix{G}}
\newcommand{\parMat}{\genericMatrix{H}}

\newcommand{\topdeg}{\ensuremath{\gamma_{\operatorname{t}}}}

\newcommand{\distor}{\ensuremath{D}}

\newcommand{\Prob}{\ensuremath{\mathbb{P}}}

\newcommand{\midWtEnum}{\ensuremath{\mathcal{A}_\midbit}}
\newcommand{\WtEnumAsymp}{\ensuremath{\mathcal{A}}}

\newcommand{\cardinal}{\ensuremath{\operatorname{card}}}

\newcommand{\vdeg}{\ensuremath{\gamma_v}}
\newcommand{\cdeg}{\ensuremath{\gamma_c}}


%% file: martin_macros.tex
\theoremstyle{plain}

\newtheorem{theos}{Theorem}
\newtheorem{props}{Proposition}
\newtheorem{lems}{Lemma}
\newtheorem{cors}{Corollary}

\theoremstyle{definition}
\newtheorem{exas}{Example}
\newtheorem{algs}{Algorithm}
\newtheorem{asss}{Asumption}
\newtheorem{defns}{Definition}

\newcommand{\btheos}{\begin{theos}}
\newcommand{\etheos}{\end{theos}}
\newcommand{\bprops}{\begin{props}}
\newcommand{\eprops}{\end{props}}
\newcommand{\bdes}{\begin{defns}}
\newcommand{\edes}{\end{defns}}
\newcommand{\blems}{\begin{lems}}
\newcommand{\elems}{\end{lems}}
\newcommand{\bcors}{\begin{cors}}
\newcommand{\ecors}{\end{cors}}
\newcommand{\bexs}{\begin{exas}}
\newcommand{\eexs}{\end{exas}}
\newcommand{\balgs}{\begin{algs}}
\newcommand{\ealgs}{\end{algs}}
\newcommand{\bass}{\begin{asss}}
\newcommand{\eass}{\end{asss}}

\long\def\comment#1{}

\newcommand{\nbit}{\ensuremath{n}}
     
\newcommand{\Exs}{\ensuremath{\mathbb{E}}}

\newcommand{\rate}{\mathrm{R}}

\newcommand{\widgraph}[2]{\includegraphics[keepaspectratio,width=#1]{#2}}


%% file: MarWai06c_arxiv.bbl
\begin{thebibliography}{10}

\bibitem{Barron:2003:it}
R.~J. Barron, Brian Chen, and G.~W. Wornell.
\newblock The duality between information embedding and source coding with side
  information and some applications.
\newblock {\em IEEE Trans. Info. Theory}, 49(5):1159--1180, May 2003.

\bibitem{Chou03}
J.~Chou, S.~S. Pradhan, and K.~Ramchandran.
\newblock Turbo and trellis-based constructions for source coding with side
  information.
\newblock In {\em Data Compression Conference (DCC)}, 2003.

\bibitem{Ciliberti05b}
S.~Ciliberti, M.~M\'{e}zard, and R.~Zecchina.
\newblock Message-passing algorithms for non-linear nodes and data compression.
\newblock Technical report, November 2005.
\newblock arXiv:cond-mat/0508723.

\bibitem{cover91:book}
Thomas~M. Cover and Joy~A. Thomas.
\newblock {\em Elements of Information Theory}.
\newblock John Wiley \& Sons, Inc., New York, 1991.

\bibitem{Litsyn_and_Shevelev}
S.~Litsyn and V.~Shevelev.
\newblock On ensembles of low-density parity-check codes: asymptotic distance
  distributions.
\newblock {\em IEEE Trans. Info. Theory}, 48(4):887--908, April 2002.

\bibitem{Marcellin90}
M.~W. Marcellin and T.~R. Fischer.
\newblock Trellis coded quantization of memoryless and {G}auss-{M}arkov
  sources.
\newblock {\em IEEE Trans. Comm.}, 38(1):82--93, 1990.

\bibitem{MarWai06a}
E.~Martinian and M.~J. Wainwright.
\newblock Low density codes achieve the rate-distortion bound.
\newblock In {\em Data Compression Conference}, volume~1, page To appear, March
  2006.

\bibitem{martinian:2003:allerton}
E.~Martinian and J.~S. Yedidia.
\newblock Iterative quantization using codes on graphs.
\newblock In {\em Allerton Conference on Communication, Control, and
  Computing}, Monticello, IL, October 2003.

\bibitem{Murayama04}
T.~Murayama.
\newblock {T}houless-{A}nderson-{P}almer approach for lossy compression.
\newblock {\em Physical Review E}, 69:035105(1)--035105(4), 2004.

\bibitem{Richardson:it:2001}
T.~J. Richardson and R.~L. Urbanke.
\newblock The capacity of low-density parity-check codes under message-passing
  decoding.
\newblock {\em IEEE Trans. Info. Theory}, 47(2):599--618, February 2001.

\bibitem{Viterbi74}
A.~J. Viterbi and J.~K. Omura.
\newblock Trellis encoding of memoryless discrete-time sources with a fidelity
  criterion.
\newblock {\em IEEE Trans. Info. Theory}, {IT}-20(3):325--332, 1974.

\bibitem{wainwright:2005:isit}
M.~J. Wainwright and E.~Maneva.
\newblock Lossy source coding via message-passing and decimation over
  generalized codewords of {LDGM} codes.
\newblock In {\em Proc. Int. Symp. Info. Theory}, September 2005.

\bibitem{WynerZiv76}
A.~D. Wyner and J.~Ziv.
\newblock The rate-distortion function for source encoding with side
  information at the encoder.
\newblock {\em IEEE Trans. Info. Theory}, IT-22:1--10, January 1976.

\bibitem{Yang05}
Y.~Yang, V.~Stankovic, Z.~Xiong, and W.~Zhao.
\newblock On multiterminal source code design.
\newblock In {\em Data Compression Conference}, 2005.

\bibitem{Zamir02}
R.~Zamir, S.~Shamai (Shitz), and U.~Erez.
\newblock Nested linear/lattice codes for structured multiterminal binning.
\newblock {\em IEEE Trans. Info. Theory}, 6(48):1250--1276, 2002.

\end{thebibliography}
